\documentstyle[12pt]{article}

\bibliography{plain}
\title{\bf  Hierarchy of Chaotic Maps with an Invariant Measure}

\vspace{20mm}

\author{M. A.
Jafarizadeh$^{a,b,c}$\thanks{E-mail:jafarzadeh@ark.tabrizu.ac.ir}
, S.Behnia$^{d,e}$,
 S.Khorram$^{f}$\thanks{E-mail:skhorram@ark.tabrizu.ac.ir} and
  H.Naghshara$^{a}$ \thanks{E-mail:naghshara@ark.tabrizu.ac.ir}.\\
$^a${\small Department of Theoretical Physics and Astrophysics,
Tabriz University, Tabriz 51664, Iran.} \\ $^b${\small Institute
for Studies in Theoretical Physics and Mathematics, Teheran
19395-1795, Iran.} \\ $^c${\small Pure and Applied Science
Research Center, Tabriz 51664, Iran.} \\ $^d${\small Plasma
Physics Research Center, IAU, Teheran 14835-159, Iran.} \\ $^e $
{\small Department of Physics, IAU, Urmia, Iran.} \\ $^f$ {\small
Center for Applied Physics Research, Tabriz University, Tabriz
51664, Iran.}} \pagebreak
\begin{document}
\maketitle
\vspace{15mm}
\newpage
\begin{abstract}
We give hierarchy of one-parameter family $F(a,x)$ of maps of the
interval $[0,1]$ with an invariant measure. Using the measure, we
calculate Kolmogorov-Sinai entropy, or equivalently Lyapunov
characteristic exponent, of these maps analytically, where the
results thus obtained have been approved with numerical
simulation. In contrary to the usual one-parameter family of maps
such as logistic and tent maps, these maps do not possess period
doubling or period-n-tupling cascade bifurcation to chaos, but
they have single fixed point attractor at certain parameter
values, where they bifurcate directly to chaos without having
period-n-tupling scenario exactly at these values of parameter
whose Lyapunov characteristic exponent begins to be positive.
 {\bf PACs
numbers:05.45.Ra, 05.45.Jn, 05.45.Tp }
\end{abstract}
\pagebreak \vspace{7cm}
\section{INTRODUCTION}
In recent years chaos or more properly dynamical systems have
become an important area of research activity. One of the
landmarks in it was introduction of the concept of
Sinai-Ruelle-Bowen (SRB) measure or natural invariant
measure\cite{sinai, jakob}. This is roughly speaking a measure
that is supported on an attractor and that describe the statistics
of the long time behavior of the orbits for almost every initial
condition in the corresponding basin of attractor . This measure
can be obtained by computing the fixed density of the so called
Frobenius-Perron operator which can be viewed as a
differential-integral operator, hence, exact determination of
invariant measure of dynamical systems is rather a nontrivial
task, such that invariant measure of few dynamical systems such as
one-parameter family one-dimensional piecewise linear maps
\cite{tasaki,tasaki1} including Baker and tent maps or unimodal
maps such as logistic map for certain values of its parameter, can
be derived analytically. In most of cases only numerical
algorithms, as an example Ulam's method\cite{froy,froy1,blank} are
used for computation of fixed densities of Frobenius-Perron
operator. Here in this article we give a hierarchy of
one-parameter family $\Phi^{(1,2)}(\alpha,x)$ of maps of interval
$[0,1]$ with an invariant measure. These maps are defined as ratio
of polynomials of degree N, where we have derived analytically
their invariant measure for an arbitrary values of the parameter
$\alpha$ and every integer values of $N$. Using this measure, we
have calculated analytically, Kolmogorov-Sinai entropy or
equivalently positive Lyapunov characteristic exponent of these
maps, where the numerical simulation of  up to degree N=10 approve
the analytic calculation. Also it is shown  that these maps have
another interesting property, that is, for even values of $N$ the
$\Phi^{(1)}(\alpha,x)$($\Phi^{(2)}(\alpha,x)$) maps   have only a
fixed point attractor $x=1$($x=0$) provided that  their parameter
belongs to interval $(N,\infty)$($(0, \frac{1}{N})$) while, at
$\alpha\geq N$ $(\alpha\geq\frac{1}{N})$ they bifurcate to chaotic
regime without having any period doubling or period-n-tupling
scenario and remain chaotic for all $\alpha\in{(0,N)}$
($\alpha\in(\frac{1}{N}, \infty)$) but for odd values of N, these
maps have only fixed point attractor $x=0$ for
$\alpha\in(\frac{1}{N}, N)$, again they bifurcate to chaotic
regime at $\alpha\geq\frac{1}{N}$, and remain chaotic for
$\alpha\in(0, \frac{1}{N})$,
 finally they bifurcate at $\alpha=N$ to have $x=1$ as fixed point attractor for all
$\alpha\in(\frac{1}{N}, \infty)$(see Figures 1,2 and 3).
 The paper is organized as
follows: In section II we introduce hierarchy of family of
one-parameter maps, In Section III we show that the proposed
anzats for the invariant measure of their maps are eigenfuntion of
Ferobenios-Perron operator with largest eigenvalue $1$, for any
finite $N$. Then in section IV using this measure we calculate
kolmogorov-Sinai entropy of these maps for an arbitrary value of
parameter $\alpha$ and every integer values of $N$. In section V
we compare analytic calculation with the numerical simulation.
Paper ends with a brief conclusion.
\section{ONE-PARAMETER FAMILIES OF CHAOTIC MAPS }
The one-parameter families of chaotic maps of the interval $[0,1]$
with an invariant measure are defined as ratio of polynomials of
degree $N$ : $$
\Phi_{N}^{(1)}(x,\alpha)=\frac{\alpha^2\left(1+(-1)^N{
}_2F_1(-N,N,\frac{1}{2},x)\right)}
{(\alpha^2+1)+(\alpha^2-1)(-1)^N{ }_2F_1(-N,N,\frac{1}{2},x)}$$
 \begin{equation}
  =\frac{\alpha^2(T_N(\sqrt{x}))^{2}}{1+(\alpha^2-1)(T_N(\sqrt{x})^{2})}\quad ,
  \end{equation}
$$ \Phi_{N}^{(2)}(x,\alpha)=\frac{\alpha^2\left(1-(-1)^N{
}_2F_1(-N,N,\frac{1}{2},(1-x))\right)}
{(\beta^2+1)-(\alpha^2-1)(-1)^N{ }_2F_1(-N,N,\frac{1}{2},(1-x))}
$$
 \begin{equation}
  =\frac{\alpha^2(U_N(\sqrt{(1-x)}))^{2}}{1+(\alpha^2-1)(U_N(\sqrt{(1-x)})^{2})}\quad ,
  \end{equation}
where $N$ is an integer greater than one. Also $$
_2F_1(-N,N,\frac{1}{2},x)=(-1)^{N}\cos{(2N\arccos\sqrt{x})}=(-1)^{N}T_{2N}(\sqrt{x})
$$ is hypergeometric polynomials of degree $N$ and
$T_{N}(x)(U_{n}(x))$ are Chebyshev polynomials of type I (type
II)\cite{wang}, respectively. Obviously these map the unit
interval $[0,1]$ into itself and are related to each other through
the following relation :
\begin{equation}
 \Phi_{N}^{(1)}(x,\alpha)=g(\Phi_{N}^{(2)},g(x),\frac{1}{\alpha})
 =g \circ \Phi_{N}^{(1)}(\frac{1}{\alpha}) \circ
 g(x)\quad\quad\mbox{for even N,}
\end{equation}
and
 \begin{equation}
 \Phi_{N}^{(1)}(x,\alpha)=\Phi_{N}^{(2)}(x,\alpha)\quad\quad\mbox{for odd N,}
\end{equation}
 where $g(x)$ is the invertible map
$g(x)=g^{-1}(x)=1-x$ and the symbol $\circ$ means composition of
functions. From now on, depending on situation, we will consider
one of these maps, since, we can get all required information
concerning the other map via using the relations $(2-3)$ and
$(2-4)$ between these two maps:
 $ \Phi_{N}^{(1)}(\alpha,x)$ is (N-1)-model map, that is it has
 $(N-1)$ critical points in unit interval $[0,1]$,(see Figure 4) since its
 derivative is proportional to derivative of hypergeometric
 polynomial $_2F_1(-N,N,\frac{1}{2},x)$ which is itself a hypergeometric
 polynomial of degree $(N-1)$, hence it has
 $(N-1)$ real roots in unit interval $[0,1]$. Defining Shwarzian
 derivative\cite{dev}
${S\Phi_N(x)}$ as: $$
S\left(\Phi_N^{(1)}(x)\right)=\frac{\Phi_{N}^{(1)\prime\prime\prime}(x)}{\Phi_{N}^{(1)\prime}(x)}-\frac{3}{2}\left({\frac{\Phi_{N}^{(1)\prime\prime}(x)}{\Phi_{N}^{(1)\prime}(x)}}\right)^2=\left(\frac{\Phi_{N}^{(1)\prime\prime}(x)}{\Phi_{N}^{(1)\prime}(x)}\right)^{\prime}-\frac{1}{2}\left(\frac{\Phi_{N}^{(1)\prime\prime}(x)}{\Phi_{N}^{(1)\prime}(x)}\right)^2,
$$
 with a prime denoting a single differentiation with respect to variable $x$, one can show that:
$$
S\left(\Phi_{N}^{(1)}(x)\right)=S\left(_2F_1(-N,N,\frac{1}{2},x)\right)\leq0,
$$
 since $\frac{d}{dx}(_2F_1(-N,N,\frac{1}{2},x))$ can be written
as: $$
\frac{d}{dx}\left(_2F_1(-N,N,\frac{1}{2},x)\right)=A\prod^{N-1}_{i=1}(x-x_i),
$$
 with $0 \leq{x_1}<{x_2}<{x_3}<....<x_{N-1}\leq{1}$ then we
have: $$
S\left(_2F_1(-N,N,\frac{1}{2},x)\right)=\frac{-1}{2}\sum^{N-1}_{J=1}\frac{1}{(x-a_j)^2}-\left(\sum^{N-1}_{J=1}\frac{1}{(x-x_j)}\right)^2<{0}.
$$
 Therefore the maps $\Phi_{N}^{(1)}(x)$ have at most $N+1$
attracting periodic orbits\cite{dev}. As we will show at the end
of this section, these maps have only a single period one stable
fixed points. \\ Denoting n-composition of functions
$\Phi^{(1,2)}(x_{1},\alpha)$ by $\Phi^{(n)}$, it is
straightforward to show that the derivative of $\Phi^{(n)}$ at its
possible n periodic points of an n-cycle:
$x_{2}=\Phi^{(1,2)}(x_{1},\alpha),
 x_{3}=\Phi^{(1,2)}(x_{2},\alpha),\cdots
, x_{1}=\Phi^{(1,2)}(x_{n},\alpha)$ is
\begin{equation}
\mid\frac{d}{dx}\Phi^{(n)}\mid=\mid\frac{d}{dx}\overbrace{\left(\Phi^{(1,2)}\circ\Phi^{(1,2)}\circ\cdots\circ\Phi^{(1,2)}(x,\alpha)\right)}^{n}\mid
=\prod_{k=1}^{n}\mid\frac{N}{\alpha}(\alpha^{2}+(1-\alpha^{2})x_{k})\mid,
\end{equation}
since for $x_{k}\in [0,1]$ we have:
$$min(\alpha^{2}+(1-\alpha^{2}x_{k}))=min(1,\alpha^{2}), $$
therefore,
$$min\mid\frac{d}{dx}\Phi^{(n)}\mid=\left(\frac{N}{\alpha}min(1,\alpha^{2})\right)^{n}.$$
Hence the above expression is definitely  greater than one for $
\frac{1}{N}< \alpha < N $, that is, both maps do not have any kind
of n-cycle or periodic orbits for $ \frac{1}{N}< \alpha < N $,
actually they are ergodic for this interval of parameter. From
(2-13) it follows that $\mid\frac{d}{dx}\Phi^{(n)}\mid$ at n
periodic points of the n-cycle belonging to interval [0,1], varies
between ${(N\alpha)}^{n}$ and ${(\frac{N}{\alpha})}^{n}$ for
$\alpha<\frac{1}{N}$ and between $(\frac{N}{\alpha})^{n}$ and
$(N\alpha )^{n}$ for $\alpha>N$, respectively.\\ From the
definition of these maps, we see that for odd N, both $x=0$ and
$x=1$ belong to one of the n-cycles, while for even N, only $x=1$
belongs to one of the n-cycles of $\Phi_{N}^{(1)}(x,\alpha)$ and
$x=0$ belongs to one of the n-cycles of
$\Phi_{N}^{(2)}(x,\alpha).$\\ For
$\alpha<(\frac{1}{N})(\alpha>{N})$, the formula $(2-5)$ implies
that for those cases in which $x=0 (x=1)$ belongs to one of
n-cycles we will have $\mid\frac{d}{dx}\Phi^{(n)}\mid<1$, hence
the curve of $\Phi^{(n)}$ starts at $x=0 (x=1)$ beneath the
bisector and then crosses it at the next (previous) periodic point
with slope greater than one (see Fig. 1), since the formula
$(2-5)$ implies that the slope of fixed points increases with the
increasing (decreasing) of $\mid x_{k}\mid$, therefore at all
periodic points of n-cycles except for $x=0 (x=1)$ the slop is
greater than one that is they are unstable, this is possible only
if $x=0 (x=1)$ is the only period one fixed point of these maps.\\
Hence all n-cycles except for possible period one fixed points
$x=0$ and $x=1$ are unstable, where  for $\alpha\in
[0,\frac{1}{N}]$, the fixed point $x=0$ is stable in maps
$\Phi_{N}^{(1,2)}(x,\alpha)$(for odd N) and
$\Phi_{N}^{(2)}(x,\alpha)$ (for even N), while for $\alpha\in
[N,\infty)$ and $\Phi_{N}^{(1)}(x,\alpha)$, the $x=1$ is stable
fixed point in maps $\Phi_{N}^{(1,2)}(x,\alpha)$(for odd N).
 \\ As an example we give below some of these maps:
$$
 \phi_{2}^{(1)}=\frac
{\alpha^{2}(2x-1)^{2}}{4x(1-x)+\alpha^{2}(2x-1)^{2}}\quad, $$
\vspace{3mm} $$
\phi_{2}^{(2)}=\frac{4\alpha^{2}x(1-x)}{1+4(\alpha^{2}-1)x(1-x)}\quad,
$$ \vspace{3mm} $$
\phi_{3}^{(1)}=\phi_{3}^{(2)}=\frac{\alpha^{2}x(4x-3)^{2}}{\alpha^{2}x(4x-3)^{2}+(1-x)(4x-1)^{2}}\quad,
$$
 \vspace{3mm}
 $$
\phi_{4}^{(1)}=\frac{\alpha^{2}(1-8x(1-x))^{2}}{\alpha^2(1-8x(1-x))^{2}+16x(1-x)(1-2x)^{2}}\quad,
$$ \vspace{3mm} $$
\phi_{4}^{(2)}=\frac{16\alpha^{2}x(1-x)(1-2x)^{2}}{(1-8x+8x^{2})^{2}+16\alpha^{2}x(1-x)(1-2x)^{2}}\quad
, $$
 \vspace{3mm}
 $$
\phi_{5}^{(1)}=\phi_{5}^{(2)}=\frac{\alpha^{2}x(16x^{2}-20x+5)^{2}}{\alpha^{2}x(16x^{2}-20x+5)^{2}+(1-x)(16x^{2}-(2x-1))}\quad.
$$
 \vspace{3mm}
 Below we also introduce their conjugate or isomorphic
maps which will be very useful in derivation of their invariant
measure and calculation of their KS-entropy in the next section.
Conjugacy means that the invertible map $ h(x)=\frac{1-x}{x} $
maps $ I=[0,1]$ into $ [0,\infty) $ and transforms maps
$\Phi_{N}^{(1,2)}(x,\alpha) $ into
$\tilde{\Phi}_{N}^{(1,2)}(x,\alpha)$ defined  as:
\begin{equation}
\left\{\begin{array}{l}
\tilde{\Phi}_{N}^{(1)}(x,\alpha)=h\circ\Phi_{N}^{(1)}(x,\alpha)\circ
h^{(-1)}=\frac{1}{\alpha^{2}}\tan^{2}(N\arctan\sqrt{x}),\\
\tilde{\Phi}_{N}^{(2)}(x,\alpha)=h\circ\Phi_{N}^{(2)}(x,\alpha)\circ
h^{-1}=\frac{1}{\alpha^{2}}\cot^{2}(N\arctan\frac{1}{\sqrt{x}}).
\end{array}\right.
\end{equation}
\vspace{5mm}
\section{INVARIANT MEASURE }
\setcounter{equation}{0}
 Dynamical systems, even apparently simple
dynamical systems as those described by maps of an interval can
display a rich variety of different asymptotic behavior. On
measure theoretical level these types of behavior are described by
SRB \cite{sinai} or invariant measure describing statistically
stationary states of the system. The probability measure $\mu$  on
$[0,1]$ is called an SRB or invariant measure of the maps
$\Phi_{N}^{(1,2)}(x,\alpha)$ given in $(2-1)$ and $(2-2)$, if it
is $\Phi_N^{(1,2)}(x,\alpha)$-Invariant and absolutely continuous
with respect to Lebesgue measure. For deterministic system such as
$\Phi_{N}^{(1,2)}(x,\alpha)$-map, the
$\Phi_N^{(1,2)}(x,\alpha)$-invariance means that its invariant
measure $\mu(x)$ fulfills the following formal Ferbenius-Perron
integral equation
$$\mu(y)=\int_{0}^{1}\delta(y-\Phi_N^{(1,2)}(x,\alpha))\mu(x)dx.$$
This is equivalent to:
\begin{equation}
\mu(y)=\sum_{x\in\Phi_{N}^{-1(1,2)}(y,\alpha)}\mu(x)\frac{dx}{dy}\quad,
\end{equation}
defining the action of standard Ferobenius-Perron operator for the
map $\Phi_N(x)$ over a function as:
\begin{equation}
P_{\Phi_{N}^{(1,2)}}f(y)=\sum_{x\in
\Phi_{N}^{-1(1,2)}(y,\alpha)}f(x)\frac{dx}{dy}\quad.
\end{equation}
We see that, the invariant measure $\mu(x)$ is given as the
eigenstate of the Frobenios-Perron operator $P_{\Phi_N^{(1,2)}}$
corresponding to largest eigenvalue 1.\\ As we will prove below
the measure $\mu_{\Phi_N^{(1,2)}(x,\alpha)}(x,\beta)$ defined as:
\begin{equation}
\frac{1}{\pi}\frac{\sqrt{\beta}}{\sqrt{x(1-x)}(\beta+(1-\beta)x)}\quad,
\end{equation}
with $\beta>0$ is the invariant measure of the maps
$\Phi_N^{(1,2)}(x,\alpha)$ provided that we choose the parameter
$\alpha$ in the following form :
\begin{equation}
\alpha=\frac{\sum{^{[\frac{(N-1)}{2}]}_{k=0}
C^{N}_{2k+1}\beta^{-k}}}{\sum{^{[\frac{N}{2}]}_{k=0}
C^{N}_{2k}\beta^{-k}}}\quad,
\end{equation}
in $\Phi_N^{(1,2)}(x,\alpha)$  maps for odd values of N and in
$\Phi_N^{(1)}(x,\alpha)$maps for even values of N
\begin{equation}
\alpha=\frac{\beta\sum{^{[\frac{(N)}{2}]}_{k=0}
C^{N}_{2k}\beta^{-k}}}{\sum{^{[\frac{(N-1)}{2}]}_{k=0}
C^{N}_{2k+1}\beta^{-k}}}\quad,
\end{equation}
in $\Phi_N^{(2)}(x,\alpha)$ maps for even values of N, where
$[\quad]$ means greatest integer part.\\
 As we see the above measure is defined only for $\beta > 0$ hence, from the
 relations $(3-4)$ and $(3-5)$, it follows that the maps $\Phi_N^{(1,2)}(x,\alpha)$ have
invariant  measure only for $\alpha\in{(\frac{1}{N}, N)}$ for odd
values of N, $\Phi_N^{(1)}(x,\alpha)$ maps have invariant measure
for $\alpha\in{(0 , N)}$ and $\Phi_N^{(2)}(x,\alpha)$ have for
$\alpha\in{(\frac{1}{N},  \infty)}$ for even N, respectively. For
other values of $\alpha$ these maps have single attractive fixed
points, which is the same as the prediction of the previous
section.
\\ In order to prove that measure $(3-3)$ satisfies equation $(3-1)$, with $\alpha $
given by relations $(3-4)$ and $(3-5)$, it is rather convenient to
consider the conjugate map:
\begin{equation}
\tilde{\Phi}_{N}^{(1)}(x,\alpha)=\frac{1}{\alpha^2}\tan^{2}(N\arctan\sqrt{x}),
\end{equation}
with measure $\tilde{\mu}_{\tilde{\Phi}_{N}^{(1)}}$related to the
measure $\mu_{\tilde{\Phi}_{N}^{(1)}}$ to the following relation:
$$\tilde{\mu}_{\tilde{\Phi}_{N}^{(1)}}(x)=\frac{1}{(1+x)^2}\mu_{\Phi_{N}^{(1)}}(\frac{1}{1+x}).$$
Denoting $\tilde{\Phi}_{N}^{(1)}(x,\alpha)$ on the left hand side
of $(3-6)$ by $y$ and inverting it, we get :
\begin{equation}
x_k=\tan^2(\frac{1}{N}\arctan\sqrt{y\alpha^2}+\frac{k\pi}{N})\quad\quad
k=1,..,N.
\end{equation}
 Then, taking derivative of $x_k$ with respect to $y$, we obtain:
\begin{equation}
\mid\frac{dx_{k}}{dy}\mid=\frac{\alpha}{N}\sqrt{x_{k}(1+x_k)}\frac{1}{\sqrt{y}(1+\alpha^2y)}\quad.
\end{equation}
Substituting the above result in equation $(3-1)$, we have:
\begin{equation}
\tilde{\mu}_{\tilde{\Phi}_{N}^{(1)}}(y)\sqrt{y}(1+\alpha^2y)=\frac{1}{N}\sum_{k}\sqrt{x_{k}}(1+x_{k}\tilde{\mu}_{\tilde{\Phi}_{N}^{(1)}}(x_k))\quad,
\end{equation}
considering the following anzats for the invariant measure
$\tilde{\mu}_{\tilde{\Phi}_{N}^{(1)}}(y)$:
\begin{equation}
\tilde{\mu}_{\tilde{\Phi}_{N}^{(1)}}(y)=\frac{1}{\sqrt{y}(1+\beta
y)}\quad,
\end{equation}
the above equation reduces to: $$\frac{1+\alpha^2y}{1+\beta
y}=\frac{\alpha}{N}\sum_{k=1}^{N}\left(\frac{1+x_{k}}{1+\beta
x_{k}}\right)$$ which can be written as:
\begin{equation}
\frac{1+\alpha^2y}{1+\beta
y}=\frac{\alpha}{\beta}+\left(\frac{\beta-1}{\beta^{2}}\right)\frac{\partial}{\partial\beta^{-1}}(\ln(\Pi_{k=1}^{N}(\beta^{-1}+x_{k}))).
\end{equation}
To evaluate the second term in the right hand side of above
formulas we can write the equation in the following form: $$
0=\alpha^{2}y\cos^{2}(N\arctan\sqrt{x})-\sin^{2}(N\arctan\sqrt{x})
$$ $$
={\frac{(-1)^{N}}{(1+x)^{N}}}\left(\alpha^{2}y(\sum_{k=0}^{[\frac{N}{2}]}C_{2k}^{N}(-1)^{N}x^{k})^{2}-x(\sum_{k=0}^{[\frac{N-1}{2}]}C_{2k+1}^{N}(-1)^{N}x^{k})^{2}\right),
$$
 $$=\frac{\mbox{constant}}{(1+x)^{N}}\prod_{k=1}^{N}(x-x_{k})\quad,
$$
 where $x_{k}$ are the roots of equation $(3-6)$, and are given by
formula $(3-7)$.Therefore, we have:
$$\frac{\partial}{\partial\beta^{-1}}\ln\left(\prod_{k=1}^{N}(\beta^{-1}+x_{k})\right)$$
 $$=\frac{\partial}{\partial\beta^{-1}}\ln\left[(1-\beta^{-1})^{N}(\alpha^{2}y\cos^{2}(N\arctan\sqrt{-\beta^{-1}})-\sin^{2}(N\arctan\sqrt{-\beta^{-1}}))\right]$$
\begin{equation}
=-\frac{N\beta}{\beta-1}+\frac{\beta
N(1+\alpha^2y)A(\frac{1}{\beta})
}{(A(\frac{1}{\beta}))^{2}\beta^{2}y+(B(
\frac{1}{\beta}))^2}\quad,
\end{equation}
with polynomials $ A(x)$ and $B(x)$ defined as: $$
 A(x)=\Pi_{k=0}^{[ \frac{N}{2}]}C_{2k}^{N}x^{k},
$$
\begin{equation}
B(x)=\Pi_{k=0}^{[ \frac{N-1}{2}]}C_{2k+1}^{N}x^{k}.
\end{equation}
In deriving the of above formula we have used the following
identities:
$$\cos(N\arctan\sqrt{x})=\frac{A(-x)}{(1+x)^{\frac{N}{2}}}\quad,
$$
\begin{equation}
\sin(N\arctan\sqrt{x})=\sqrt{x}\frac{B(-x)}{(1+x)^{\frac{N}{2}}}\quad,
\end{equation}
 inserting the results $(3-12)$, in $(3-6)$, we get:
$$\frac{1+\alpha^{2}y}{1+\beta y}=\frac{1+\alpha^{2}y}{\left(
\frac{B( \frac{1}{\beta})}{\alpha A( \frac{1}{\beta})}+\beta(
\frac{\alpha A( \frac{1}{\beta})}{B(
\frac{1}{\beta})}\right)}\quad.$$
 Hence to get the final result we have to choose the parameter
 $\alpha$ as:
$$\alpha =\frac{B( \frac{1}{\beta})}{A(\frac{1}{\beta})}\quad.$$
 With the procedure similar to the one given above we could
 get the relation $(3-11)$ between the parameters $\alpha$ and $\beta$
for the second kind of maps.
\\
\section{KOLMOGROV-SINAI ENTROPY}
\setcounter{equation}{0} Kolomogrov-Sinai entropy (KS) or metric
entropy measure \cite{sinai}how chaotic a dynamical system is and
it is proportional to the rate at which information about the
state of dynamical system is lost in the course of time or
iteration. Therefore, it can also be defined as the average rate
of loss information for a discrete measurable dynamical system
$(\Phi_N^{(1,2)}(x,\alpha),\mu)$, by introducing a partition
$\alpha={A_c} (n_1,.....n_{\gamma})$ of the interval $[0,1]$ into
individual laps $A_i$ one can define the usual entropy associated
with the partition by:
$$H(\mu,\gamma)=-\sum^{n(\gamma)}_{i=1}m(A_c)\ln{m(A_c)},$$ where
$m(A_c)=\int{_{n\in{A_i}}\mu(x)dx}$ is the invariant measure of $
A_i$. Defining n-th refining $\gamma(n)$ of  $ \gamma$:
$$\gamma^{n}=\bigcup^{n-1}_{k=0}(\Phi_N^{(1,2)}(x,\alpha))^{-(k)}(\gamma),$$
and defining an entropy per unit step of refining by :
$$h(\mu,\Phi_N^{(1,2)}(x,\alpha),\gamma)=\lim{_{n\rightarrow{\infty}}}(\frac{1}{n}H(\mu,\gamma)),$$
if the size of individual laps of $\gamma(N)$ tends to zero as n
increases, then the above entropy is known as Kolmogorov-Sinai
entropy, that is:
$$h(\mu,\Phi_N^{(1,2)}(x,\alpha))=h(\mu,\Phi_N^{(1,2)}(x,\alpha),\gamma).$$
KS-entropy , which is a quantitative measure of the rate of
information lost with the refining, may also be written as:
\begin{equation}
h(\mu,\Phi_N^{(1,2)}(x,\alpha))=\int{\mu(x)dx}\ln{\mid\frac{d}{dx}\Phi_N^{(1,2)}(x,\alpha)\mid},
\end{equation}
which is also a statistical mechanical expression for the Lyapunov
characteristic exponent, that is, mean divergence rate of two
nearby orbits. The measurable dynamical system
$(\Phi_N^{(1,2)}(x,\alpha),\mu)$ is chaotic for $h>0$ and
predictive for $h=0$. \\ In order to calculate the KS-entropy of
the maps $\Phi_N^{(1,2)}(x,\alpha)$, it is rather convenient to
consider their conjugate maps given by $(2-6)$, since it can be
shown that KS-entropy is a kind of topological invariant, that is,
it is preserved under conjugacy map, hence we have:
\begin{equation}
h(\mu,\Phi_N^{(1,2)}(x,\alpha))=h(\tilde{\mu},\tilde{\Phi}_N^{(1,2)}(x,\alpha)).
\end{equation}
Using the integral $(4-1)$, the KS-entropy of
$\Phi_{N}^{(2)}(\alpha,x)$ can be writen as\\ $$
h(\mu,\Phi_{N}^{(2)}(x,\alpha))=h(\tilde{\mu},\tilde{\Phi}_{N}^{(2)}(x,\alpha))
$$
$$=\frac{1}{\pi}\int_{0}^{\infty}\frac{\sqrt{\beta}dx}{\sqrt{x}(1+\beta
x)}\ln(\mid\frac{1}{a^{2}}\frac{d}{dx}(\cot^{2}(N
\arctan\sqrt{x}))\mid)$$\\
\begin{equation}
=\frac{1}{\pi}\int_{0}^{\infty}\frac{\beta dx}{\sqrt{x}(1+\beta
x)}ln\left(\frac{N}{\alpha^{2}}\times\frac{1}{\sqrt{x}(1+x)}\frac{\cos
N(\arctan\sqrt{x})}{\sin^{3}N(\arctan\sqrt{x})}\right).
\end{equation}
Using the relations given in $(3-14)$ we have
\begin{equation}
h(\mu,\Phi_{N}^{(2)}(x,\alpha))
=\frac{1}{\pi}\int_{0}^{\infty}\frac{\sqrt{\beta}dx}{\sqrt{x}(1+\beta
x)}\ln\left(
\frac{N}{\alpha^{2}}\mid\frac{(1+x)^{N-1}A(-x)}{x^{2}(B(-x))^{3}}\mid\right),
\end{equation}
for even N, and \\
\begin{equation}
h(\mu,\Phi_{N}^{(2)}(x,\alpha))
=\frac{1}{\pi}\int_{0}^{\infty}\frac{\sqrt{\beta}dx}{\sqrt{x}(1+\beta
x)}\ln\left(\frac{N}{\alpha^{2}}\mid\frac{(1+x)^{N-1}B(-x)}{(A(-x))^{3}
}\mid\right).
\end{equation}
for odd N.\\ Considering again the relations given in $(3-14)$, we
see that polynomials appearing in the numerator ( denominator ) of
integrand appearing on the right hand side of equation $(4-5)$,
have $\frac{[N-1]}{2}$ $(\frac{[N]}{2})$ simple roots, denoted by
 $
  x_{k}^{B}\quad k=1,...,[\frac{N-1}{2}]\quad$
$(x_{k}^{A}\quad k=1,...,[\frac{N}{2}])
$
 in the interval $[0,\infty)$.
 Hence, we can write the above formula in the following form:
$$h(\mu,\Phi_{N}^{(2)}(x,\alpha))
=\frac{1}{\pi}\int_{0}^{\infty}\frac{\sqrt{\beta}dx}{\sqrt{x}(1+\beta
x)}\ln\left(\frac{N}{\alpha^{2}}\times\frac{(1+x)^{N-1}\prod_{k=1}^{[\frac{N}{2}]}\mid
x-x_{k}^{A}\mid}{x^{2}\prod_{k=1}^{[\frac{N-1}{2}]}\mid
 x-x_{k}^{B}\mid^{3}}\right),$$
for even N, and $$h(\mu,\Phi_{N}^{(2)}(x,\alpha))
=\frac{1}{\pi}\int_{0}^{\infty}\frac{\sqrt{\beta}dx}{\sqrt{x}(1+\beta
x)}\ln\left(\frac{N}{\alpha^{2}}\times\frac{(1+x)^{N-1}\prod_{k=1}^{[\frac{N-1}{2}]}\mid
x-x_{k}^{B}\mid}{\prod_{k=1}^{[\frac{N}{2}]}\mid
x-x_{k}^{A}\mid^{3}}\right).$$ for odd N. \\ Now making the
following change of variable
$x=\frac{1}{\beta}\tan^{2}\frac{\Theta}{2}$, and taking into
account that degree of numerators and denominator are equal for
both even and odd values of N, we get
$$h(\alpha,\Phi_{N}^{(2)}(x,\alpha))=\frac{1}{\pi}\int_{0}^{\infty}d\theta\{\ln(\frac{N}{\alpha^{2}})+(N-1)\ln\mid(\beta+1+(\beta-1)\cos\theta\mid)$$
$$
+\sum_{k=1}^{[\frac{N}{2}]}\ln\mid1-x_{k}^{A}\beta+(1+x_{k}^{A}\beta)\cos\theta\mid-3\sum_{k=1}^{[\frac{N-1}{2}]}\ln\mid1-x_{k}^{B}\beta+(1+x_{k}^{B}\beta)\cos\theta\mid
$$
 $$-2\ln\mid 1+cos\theta\mid \},$$
for even and
$$h(\alpha,\Phi_{N}^{(2)}(x,\alpha))=\frac{1}{\pi}\int_{0}^{\infty}d\theta\{\ln(\frac{N}{\alpha^{2}})+(N-1)\ln\mid(\beta+1+(\beta-1)\cos\theta\mid)
$$ $$ +\sum_{k=1}^{[\frac{N-1}{2}]}\ln
\mid1-x_{k}^{B}\beta+(1+x_{k}^{B}\beta)\cos\theta\mid
-3\sum_{k=1}^{[\frac{N}{2}]}\ln\mid
1-x_{k}^{A}\beta+(1+x_{k}^{A}\beta)\cos\theta\mid\}.$$ for odd
N.\\
 Using the following integrals:
$$ \frac{1}{\pi}\int_{0}^{\pi}\ln\mid a+b\cos\theta\mid= \left\{
\begin{array}{l} \ln\mid\frac{a+\sqrt{a^{2}-b^{2}}}{2}\mid\quad\quad\mid
a\mid >\mid b\mid
\\ \ln\mid\frac{b}{2}\mid\quad\quad\quad\quad\mid
a\mid \leq\mid b\mid,
\end{array}\right.$$
 we get
$$ h(\alpha,\Phi_{N}^{(2)}(x,\alpha))=\left\{
\begin{array}{l} \ln
\left(\frac{N}{\alpha^{2}}\frac{(\beta+1+2\sqrt{\beta})^{N-1}\prod_{k=1}^{[\frac{N}{2}]}(1+x_{k}^{A}\beta)}{\left(\prod_{k=1}^{[\frac{N-1}{2}]}(1+x_{k}^{B}\beta)\right)^{3}}\right)\quad\quad\quad\mbox{for
even N}
\\ \\ \ln\left(\frac{N}{\alpha^{2}}\frac{(\beta+1+2\sqrt{\beta})^{N-1}\prod_{k=1}^{[\frac{N-1}{2}]}(1+x_{k}^{B}\beta)}{\left(\prod_{k=1}^{[\frac{N}{2}]}(1+x_{k}^{A}\beta)\right)^{3}}\right)\quad\quad\quad\mbox{for
odd N},
\end{array}\right.$$
or $$h(\alpha,\Phi_{N}^{(2)}(x,\alpha))=\left\{
\begin{array}{l} \ln
\left(\frac{N}{\alpha^{2}}\frac{(1+\beta+2\sqrt{\beta})^{N-1}\beta^{2}A(\frac{1}{\beta})}{\beta^{(N-1)}\left(B(\frac{1}{\beta})\right)^{3}}\right)\quad\quad\quad\mbox{for
even N}
\\ \\ \ln\left(\frac{N}{\alpha^{2}}\frac{(1+\beta+2\sqrt{\beta})^{N-1}B(\frac{1}{\beta})}{\beta^{(N-1)}\left(A(\frac{1}{\beta})\right)^{3}}\right)\quad\quad\quad\quad\mbox{for
odd  N}.
\end{array}\right.$$
Using the relation:
$$\alpha=\left\{
\begin{array}{l} \beta
\frac{A(\frac{1}{\beta})}{B(\frac{1}{\beta})}\quad\quad\quad\quad\mbox{for
even N}
\\ \\
\frac{B(\frac{1}{\beta})}{A(\frac{1}{\beta})}\quad\quad\quad\quad\mbox{for
odd  N},
\end{array}\right.$$
we get
\begin{equation}
 h(\mu,\Phi_{N}^{(2)}(x,\alpha))=\ln\left(\frac{N(1+\beta+2\sqrt{\beta})^{N-1}}{(\Pi_{k=0}^{[ \frac{N}{2}]}C_{2k}^{N}\beta^{k})(\Pi_{k=0}^{[
 \frac{N-1}{2}]}C_{2k+1}^{N}\beta^{k})}\right).
\end{equation}
With a calculation rather similar to the one given above we can
calculate the KS-entropy of the $\Phi_N^{1}(x,\alpha)$ maps where
the results are the same as those given by $(4-6)$.\\ The
KS-entropy $(4-6)$ is invariant with respect to
$\beta\longrightarrow(\frac{1}{\beta})$, therefore, it has the
same asymptotic behavior near $\beta\longrightarrow 0$ and
$\beta\longrightarrow\infty$ where its asymptotic forms read
$$\left\{ \begin{array}{l}
h(\mu,\Phi_{N}^{(1,2)}(x,\alpha=N+0^{-}))\sim
(N-\alpha)^{\frac{1}{2}}\\
h(\mu,\Phi_{N}^{(1,2)}(x,\alpha=\frac{1}{N}+0^{+}))\sim
(\alpha-\frac{1}{N})^{\frac{1}{2}},
\end{array}\right.$$
for odd N and:
$$\left\{
\begin{array}{l}
h(\mu,\Phi_{N}^{(1)}(x,\alpha=N+0^{-}))\sim
(N-\alpha)^{\frac{1}{2}}\\
h(\mu,\Phi_{N}^{(2)}(x,\alpha=\frac{1}{N}+0^{+}))\sim
(\alpha-\frac{1}{N})^{\frac{1}{2}},
\end{array}\right.$$
for even N. The above asymptotic form indicates that the maps,
 $\Phi_{N}^{(1,2)}(x,\alpha)$ belong to the same universality class
which is different from the universality class of pitch fork
bifurcating maps but their asymptotic behavior is similar to class
of intermittent maps\cite{pomeau, hirsch}, even though
intermittency can not occur in these maps for any values of
parameter $\alpha$, since the maps $\Phi_{N}^{(1,2)}(x,\alpha)$
and their n-composition $\Phi^{(n)}$ do not have minimum values
other than zero and maximum values other than one in $[0,1].$
\section{SIMULATION}
\setcounter{equation}{0} Here in this section we try to calculate
Lyapunov characteristic exponent of maps $
\Phi_{N}^{(1,2)}(x,\alpha)$, $N=1,2,....,5$ in order to
investigate these maps numerically. In fact, Lyapunov
characteristic exponent is the characteristic exponent of the rate
of average magnificent of the neighborhood of an arbitrary point
$x_{0}$  and it is denoted by $ \Lambda(x_{0}) $ which is written
as: $$
\Lambda^{(1,2)}(x_{0})=lim_{n\rightarrow\infty}\ln(\mid\overbrace{\Phi_{N}^{(1,2)}(x,\alpha)
\circ \Phi_{N}^{(1,2)} ....\circ
\Phi_{N}^{(1,2)}(x_{K},\alpha)}^{n}\mid $$
\begin{equation}
=lim_{n\rightarrow\infty}\sum_{k=0}^{n-1}\ln\mid\frac{d\Phi_{N}^{(1,2)}(x_{k},\alpha)}{dx}\mid,
\end{equation}
where $ x_{k}=\overbrace{\Phi_{N}^{(1,2)} \circ \Phi_{N}^{(1,2)}
\circ ....\circ \Phi_{N}^{(1,2)}(x_{0})}^{k} $ . It is obvious
that $\Lambda^{(1,2)}(x_0)<0 $ for an attractor,
$\Lambda^{(1,2)}(x_{0})>0$ for a repeller and
$\Lambda^{(1,2)}(x_{0})=0$ for marginal situation. Also the
Liapunov number is independent of initial point $x_{0}$, provided
that the motion inside the invariant manifold is ergodic, thus
$\Lambda^{(1,2)}(x_{0})$ characterizes the invariant manifold of
$\Phi_{N}^{(1,2)}$ as a whole. For values of parameter $\alpha$ or
$\beta$, such that the map $\Phi_{N}^{(1,2)}$ be measurable,
Birkohf ergodic  theorem implies equality of KS-entropy and
Liapunov characteristic exponent, that is:
\begin{equation}
h(\mu,\Phi_{n}^{(1,2)})=\Lambda^{(1,2)}(x_{0},\Phi_{N}^{(1,2)}),
\end{equation}
Comparison of analytically calculated KS-entropy of maps
$\Phi_{N}^{(1,2)}(x,\alpha)$ for $N=1,2,\cdots \ 10$ , (see
Figures $5,6$ and $7$ for $N=2$ and $3$ ) with the corresponding
Lyapunov characteristic exponent obtained by simulation, indicate
that in chaotic region, these maps are ergodic as Birkohf ergodic
Theorem predicts. In non chaotic region of parameter Lyapunov
characteristic exponent is negative, since in this region we have
only stable period one fixed points without bifurcation. In
summation, combining the analytic discussion of section II  with
the numerical simulation we deduce that these maps are ergodic in
certain values of their parameter as explained above and in
complementary interval of parameter they have only a single period
one attractive fixed point, such that in contrary to the most of
usual one-dimensional one-parameter family of maps they have only
a bifurcation from a period one attractive fixed point to chaotic
state or vise versa.\\
\section{Conclusion}
\setcounter{equation}{0}
 We have given hierarchy of exactly
solvable one-parameter family of one-dimensional chaotic maps with
an invariant measure, that is measurable dynamical system with an
interesting property of being either chaotic (proper to say
ergodic ) or having  stable period one fixed point and they
bifurcate from a stable single periodic state to chaotic one and
vice-versa without having usual period doubling or
period-n-tupling scenario.
\\ Perhaps this interesting property is due to existence of
invariant measure for a range of values of parameter of these
maps. Hence, to approve this conjecture, it would be interesting
to find other measurable one parameter maps, specially higher
dimensional maps, which is under investigation.

\end{document}